\documentstyle[aps,eqsecnum,twocolumn,prb]{revtex}
\begin{document}
\draft
\title{Grain boundary component in W-Ga composites: a way towards skeleton
 structures}
\author{W. Krauss and H. Gleiter$^*$}
\address{Technical Physics, University of Saarbr\"ucken, D-66041
Saarbr\"ucken,
 Germany
 $^*$Forschungszentrum Karlsruhe GmbH, Postfach 3640, D-76021 Karlsruhe
}
\date{\today}
\maketitle

\begin{abstract}
Nanostructured materials consist of crystalline and grain boundary components. In the 
simplest case, both components are chemically identical. Here, we present the results of a study of a
system consisting of a crystalline component built by A atoms (tungsten) and a grain boundary 
component of B atoms (gallium). Within this system, component B is in a disordered state.
Most likely it exhibits an 'amorphous-like' structure, and coats the tungsten crystals 
uniformily with a constant thickness, thus forming a Ga skeleton structure. The non-crystalline gallium 
seems to undergo no first order structural phase transitions, e.g., no first order melting transition was
noted when the composit was below, at or above the equilibrium melting point of Ga. 
The properties of gallium as grain boundary component differ significantly from those of crystalline and
amorphous bulk gallium.
\end{abstract}
\vspace{1cm}

%
\section{Introduction}
Technical materials like, e.g., metals or ceramics, are often used as polycrstalline materials. 
Usually, the fraction of atoms within grain boundaries is many orders of magnitude  smaller than the 
fraction of 
atoms within individual grains. In contrast, the invention of nanocrystalline materials in 
1981 \cite{Gle81} has yielded the possibility to synthesize solids with nm size grains 
so that the number of atoms located within grain boundaries and grains are comparable. 
This leads to an enhanced interface surface-to-volume ratio, and hence 
mechanical, electrical, magnetical or thermal properties are affected significantly by the atoms within 
the grain boundaries \cite{Sie90,Gle94}.

Beside generating nanocrystalline materials consisting of only one kind of atoms, it should be also 
possible to produce nanocrystalline materials in which the crystallites and the interfaces have different
chemical compositions (e.g., A and B) [c.f. Fig. \ref{material}]. Materials suitable to generate nanocrystalline solids of this type
have to meet the following two conditions. First, they should be unmiscible. 
Furthermore, the material used for the grain boundary component should wet the 
material used for the formation of the crystalline component. One way of synthesis is to 
produce small crystallites of A and coat them with B. By subsequently
 compacting the 
coated particles one obtains a composite consisting of nanometer-sized grains of A atoms, separated 
by a grain boundary 
component of B atoms. The new aspect of such skeleton structures is, that it can be 
expected that the B atoms within the grain boundaries are prevented to crystallize in their equilibrium 
structure because of the interatomic forces excerted by the adjacent crystallites of A atoms.

By investigating such a two-phase material, the major questions, we shall focus on, are:

\noindent
(i) What kind of structure do the B atoms form within the grain boundaries?

\noindent
(ii) Does the grain boundary component of B atoms undergo structural phase transitions upon heating/cooling?

\noindent
(iii) What can be learned about thermal excitations and electronic properties of the boundary component?\\
\indent
In the literature, some attempts are reported to synthesize such structures. Although some 
experimental results \cite{Jan90,Ogu78,Hau92} exist and are confirmed by theoretical 
considerations \cite{Tur90,Yan93}, it was not yet to manufacture pure skeleton structures 
\cite{Boo92,Kon97}, where component B just forms the grain boundary component without larger agglomerates.
The experiments and the model, with resembles the thermodynamics and the
processing 
mechanism presented here, are graphite encapsulated nanocrystalline particles
\cite{Hor98,Ell98}.

In this publication, we present measurements of structural, thermal, electrical and magnetic 
properties of W-Ga composites. Peculiarities of physical properties observed for these 
samples will be discussed with respect to the gallium grain boundary component (gallium skeleton).

\section{Experimental}
Tungsten (W) as a high refractory metal and gallium (Ga) as a low-melting metal (T$_m$ = 29$^\circ$C)
 seem to be a suitable
model system for the formation of two-component composites of the type described in Fig. \ref{material}.
 The wetting properties of Ga on 
W surfaces is well known from FIM experiments \cite{Nis73,Kon81}. Furthermore, the 
solubility of W in Ga even at temperatures as high as 815$^{\circ}$C
 is 0.001 to 0.008 wt$\%$ 
\cite{Mas86}. The nanocrystalline samples were produced by a modified inert gas condensation method 
\cite{Hah90,Haa92}. The high refractory W was dc-sputtered from a W target (purity 99.99 
$\%$), while the Ga (purity 99.9999 $\%$) was evaporated using a conventional evaporation 
source from a W boat. The inert gas atmosphere consisted of Ar (p$_{Ar}$ $\approx$ 7 
mbar). The experimental setup is shown in Fig. \ref{inert-gas}. A variation of the process 
parameters like, e.g., inert gas pressure, power or source-collector distance can significantly 
influence the particle formation \cite{Hah90,Haa92}. The formation of nanocrystalline W crystals takes 
place within the Ar atmosphere, while the evaporated Ga coats these crystallites subsequently. These 
crystallites are then transferred due to the gas convection between the hot source and the cold collecting device
to a cooling finger, which is held at liquid nitrogen temperature. The nm-sized crystallites 
collected on the surface of the cooling finger were subsequently stripped off and were then compacted (p$_{c}$ 
$>$ 1 GPa) to macroscopic samples of typically 8 mm in diameter and thicknesses of 0.3 to 1 
mm under UHV conditions. The thickness of the Ga layer
depends on the Ga vapor pressure within the vacuum chamber. The Ga layer thicknesses were varied from 
less than a monolayer Ga up to five monolayers. 
The thicknesses of the layers were determined by measuring the overall amount of Ga within the 
samples as well as the grain size \cite{Wei93}.

\section{Results and discussion}
Pure nanocrystalline W specimen were synthesized as reference samples. For the pure W as well as for W-Ga samples,
the grain sizes of the W crystallites were in the range of 8 to 13 nm. Figure 
\ref{XRD(W,W/Ga)}(a) shows the diffraction pattern (wide angle X-ray diffraction, XRD, Cu K$_{\alpha}$ 
radiation) of the reference sample, having a grain size of $\approx$ 13 nm. The diffraction pattern indicates that
 the samples do not have the equilibrium bcc structure of W \cite{Kra97}. In fact, the W crystallized in a 
metastable modification, which is known as A15 or $\beta$-W structure \cite{Cor93}. 
Figure \ref{XRD(W,W/Ga)} (b) displays an XRD measurement of a nanocrystalline W-Ga sample with a Ga 
content of $\approx$ 35 at$\%$. Even for Ga contents as high as 45 at$\%$, the XRD pattern displayed only
 the maxima which correspond to A15-W. Bragg 
peaks due to Ga crystals were not observed, indicating that little or none of the 
Ga incorporated within the samples is crystalline. Moreover, neither a 
Laue background nor additional broad maxima, which would be characteristic for a disordered or 
ordered W-Ga solid solution, could be detected. The structural analysis of the W crystallites in the 
W-Ga composite shows, that no Ga is incorporated within the W lattice, i.e., the lattice parameters as 
well as the amount of strain is unchanged with respect to pure nanocrystalline W-A15 samples (lattice parameter
of the crystalline A15 lattice 5.05 $\AA$). 
The amount of gaseous impurities determined by hot extraction
 was equal for all samples, i.e., N$_2$ $\approx$ 2 at$\%$, H$_2$ $\approx$ 1.5
at$\%$ and O$_2$ $\approx$ 2.5 at$\%$.

Let us now address to the major question, where is the Ga located 
 and what is the structure of the Ga? There seem to be three conceivable reasons, why no Ga diffraction 
effects were observed:

\noindent
(a) The Ga atoms are agglomerated, forming very small particles of a second component, 
which cannot be detected by XRD measurements by means of line broadening.

\noindent
(b) The Ga atoms coat the nanocrystalline W grains and form an epitactic layer, which also has an A15 
structure like the W nanocrystals.

{\noindent
(c) The W-Ga composite represents an ideal system with respect to the case that the W-A15 
particles are the crystalline component, whereas the Ga forms a skeleton, which structure corresponds to none of the known Ga
modifications. The Ga skeleton may show an amorphous X-ray diffraction signal, which probably 
cannot be distinguished from the W-diffraction effects because of the large difference in the electron densities for Ga and W, 
respectively.}

To clearify, which model should be  adapted, thermal expansion, DSC, TEM, PAC, resistivity measurements as well as 
the magnetic behavior of W-Ga composites were investigated. Generally, if the thermal expansion 
of a nanocrystalline material consisting of one sort of atoms is explored, a deviation from the linear expansion at temperatures 
above room temperature has been observed \cite{Kli92}. This result is interpreted in terms of
 relaxation processes in the grain boundaries of the nanocrystalline materials. This type of reduction of the thermal
  expansion was also observed for pure nanocrystalline W-A15 samples [c.f. Fig. \ref{DSC}(a)]. However,
 the thermal expansion of nanocrystalline W-Ga composites differs significantly: No relaxation above room temperature
 was noticed [c.f. Fig. \ref{DSC}(b)]. This points to the fact that a fundamental 
difference between grain boundaries in single-component nanocrystalline materials as, e.g., W and in 
 nanocrystalline composits, exists. Furthermore, no volume changes due to the melting of Ga wa recorded if the W-Ga specimen
 were heated to temperatures above the equilibrium melting temperature of Ga.

The absence of a solid/liquid phase transition of Ga at the thermodynamic melting point was 
analyzed in more detail by means of DSC measurements. The measurements were performed 
with an isoperibolic calorimeter in the low-temperature regime (15 K $<$ T$_{meas}$ $<$ 200 K) 
and a standard DSC (Perkin-Elmer DSC 7) in the temperature range of 200 K $<$ T$_{meas}$ $<$ 
700 K. In the entire temperature range of 15 up to 700 K no DSC signal of a  solid/liquid phase transition or 
glass transition could be detected. Figure \ref{DSC1} (a) shows the thermal expansion and the DSC signalk of W-30 at$\%$ Ga specimen
in the temperature range of 210 to 370 K. A reference sample of pure Ga of about the same amount of Ga as in a W-Ga composites 
[Fig. \ref{DSC1} (a)] revealed
clearly the melting enthalpy of Ga (Fig. \ref{DSC1}).

Furthermore, conventional transmission electron microscopy was implemented to analyze the 
samples with respect to Ga agglomerates. Irrespective of the sample preparation method (ion milling, 
microtom cutting or simply breaking), no Ga crystallites nor 
Ga agglomerates of a second phase were observed. The Ga atoms, which 
could be detected by EDX measurements during SEM
 and TEM analysis, remained invisible in dark/bright field TEM.

If all observations reported so far, are compared with the 
models discussed above, the idea of an epitactic Ga layer, which exhibits an A15 structure, as well as the model of
 an amorphous Ga layer appear to be consistent with the experiments.

An additional analysis of the structure of the Ga phase was performed by using the perturbated 
$\gamma$-$\gamma$-angle correlation (PAC) technique, which has been applied to study the structure of cold-
condensed Ga films \cite{Heu79}. Samples containig about 30 at$\%$ Ga were used for these 
investigations. $^{111}$Cd was used as probe, which was 
introduced in the nanocrystalline W-Ga samples by diffusion. Temperature and time for this process 
 were chosen such, that no structural changes occured according to XRD
 and the calorimetry measurements. For example, the grain diameter of the W-A15 crystallites remained 
unchanged. Figure \ref{PAC} displays a PAC spectrum measured at a temperature of 280 K [Fig. \ref{PAC}(a)] and 
the corresponding 
frequency analysis [Fig. \ref{PAC}(b), line]. The spectrum shows a shape which corresponds to a 
highly disturbed, non-crystalline Ga structure within the nanocrystalline W-Ga composites. For 
comparison, the frequency analysis of amorphous Ga [c.f. Fig. \ref{PAC} (b), dotted line) in terms of a continious random packing model, measured at T $\leq$ 
15 K \cite{Heu79}, is also displayed. In the entire temperature range of 77 K $\leq$ 
T$_{meas}$ $\leq$ 700 K, the PAC signals were unaffected by temperature changes and look 
like Fig. \ref{PAC} (a). Even at or above the equilibrium melting temperature of $\alpha$-Ga, the PAC 
signal showed no detectable variation. A variation was noticed, however, if grain growth occured or if the structure of the W
crystallites was changed from A15 to bcc by means of annealing. These observations seem to indicate that the boundary
conditions excerted by the W crystallites on the Ga atoms that control the atomic arrangement of the Ga \cite{Wol94}.

In order to obtain further information on the Ga structure, additional experiments on the electrical and 
magnetic behavior of the samples were performed. Figure \ref{resistivity}(b) shows a measurement of the specific electrical resistance of a pure
 W-A15 sample with an average grain diameter of 10 nm, measured by the four-point 
probe method. A high absolute value of the specific resistance and a 
positive temperature coefficient of resistivity (TRC) is noted.  W-Ga samples with a Ga content of $\approx$ 30 at$\%$ and a
grain size of 10 nm (A15 structure) [Fig. \ref{resistivity}(b)] exhibited a significant difference compared to pure 
W-A15 samples. The absolute resistivity value was lower than for pure W-A15, and the TRC was negative. 
Hence, the TRC differs not only from the one of pure W-A15 [c.f. Fig. 
\ref{resistivity}(a)], but also from the TRC of pure Ga, which is positive as well in the crystalline 
as in the liquid state \cite{Ges72,Yaq66,Buc63,Gin86}.

The changed TCR of a W-Ga composite (compared to a pure W-A15 sample with comparable grain size) may be
interpreted by assuming 
that the current only flows primarily through the grain boundary phase of Ga. Therefore, the measured 
specific resistivities have to be normalized to the volume portion of Ga present in the sample. 
Figure \ref{resistivity-ga}(a) shows a schematics of the sample, whereas Fig. \ref{resistivity-ga}
(b)-(d) show measurements of the specific resistance of W-Ga composites for different Ga 
contents. In all cases, the grain diameter of the W-A15 nanocrystals is $\approx$ 10 nm.

The data may be understood as follows. With increasing Ga content within the samples, which 
corresponds to an increasing thickness of Ga in between the W-A15 grains, the resistivity 
decreases from about 145 $\mu\Omega$cm (for $\approx$ 26 at$\%$ Ga) to 94 
$\mu$$\Omega$cm (for $\approx$ 40 at$\%$ Ga). Compared to these values, the resistivities of 
$\alpha$-Ga ($\approx$ 53 $\mu$$\Omega$cm) and liquid Ga ($\approx$ 26 $\mu$$\Omega$cm) are 
lower \cite{Kir80}. Several samples were investigated, all exhibiting a negative TRC. The appearance 
of a negative TRC of Ga is contrary to all observations up to now. Parallel to the decrease of the 
specific resistance with increasing Ga content, the slighly negative TRC decreases which, for 
samples with a Ga content of 40 at$\%$, leads to a nearly constant resistivity in the 
temperature range of 5 to 285 K, fluctuating only 0.01 $\mu\Omega$cm around its mean value of $\approx$ 94 $\mu\Omega$cm.

The steep drop of the specific resistance below 5 K, which is indicated by the arrows in Figs. 
\ref{resistivity-ga}(b)-(d), was investigated in more detail with a Foner Magnetometer (VSM). If 
this drop is caused by the onset of superconductivity, VSM measurements should exhibit the 
diamagnetic behavior of a superconducting material. No Meissner-Ochsenfeld effect was revealed for pure 
nanocrystalline W-A15 samples and W-Ga 
composites ($\approx$ 35 at$\%$ Ga) in the 
temperature range of 3.8 K $\leq$ T$_{meas}$ $\leq$ 300 K and 5 K $<$ T$_{meas}$ $\leq$ 
300 K, respectively. For temperatures lower than 5 K,  the W-Ga composite showed the 
behavior of a type-II superconductor [Fig. \ref{magnetic}]. The onset of superconductivity for temperatures below 5 
K lies between the tempeartures of the onset of superconductivity of crystalline (1.07 K) 
and of amorphous (8.4 K) Ga. The onset of superconductivity at 5 K may originate from proximity effects or from the 
presence of a new Ga structure within the composites. The considerable 
hysteresis (Fig. \ref{magnetic}) points to a strong flux pinning in a highly disturbed material \cite{Buc84}. An 
estimation  \cite{Cam64} of the Ginzburg-Landau parameter $\kappa$ of the sample shown 
in Fig. \ref{magnetic} yields $\kappa$ $\approx$ 30. This high value implies, that this sample is a 
strong type-II superconductor with a strongly reduced mean free path. For comparison, the mean free path in crystalline 
bulk Ga, as large as 2 mm at 4.2 K \cite{Coc65}. The W-Ga 
composit has kept its superconductivity below 5 K, whether superconductive properties were not affected by heating the W-Ga
composits above the melting point of bulk crystalline (T$_m$$^Ga$) Ga. This result seems to agree with the DSC measurements
reported preeviously: the DSC measurements revealed no phase transition at T$_m$$^Ga$. Apparently, the atomic arrangement
formed by the Ga atoms in the grain boundaries is fixed by the boundary conditions, 
which are determined by the adjacent nanocrystalline W crystallites.

\section{Conclusions}
W-Ga composites seem to be a suitable system with respect to the production of two-component 
composites, where the nanometer-sized grains of a first component (e.g., W) are coated by material
of a second boundary component (e.g., Ga). The compacted composites may be regarded 
as a material, having a skeleton structure, whereby the skeleton is formed by the grain 
boundary component. Structure and stability of the boundary component is dominated by the nanometer-sized crystallites. The 
results reported above indicate, that the Ga grain boundary component has an atomic structure which 
seems to differ from the crystalline as well as from the amorphous state of Ga. This result seems to suggest that the
 boundary 
conditions excerted by the W-A15 nanocrystals on the Ga atoms determine the Ga structure. These boundary
conditions appear to control the properties of the Ga boundary component. In the entire temperature range (5~$\ldots$~700 K)
studied, the grain boundary component neither exhibited a glass transition nor was melting detectable.

It could be shown that the synthesis of W-Ga composites leads to samples, which physical 
properties are strongly determined by the grain boundary skeleton. It appears likely, that other alloy systems which fulfill
 the conditions {\it 
unmiscible} and {\it wetting}, are suitable for synthesizing nanocrytsalline composite materials 
consisting of a similar skeleton structure, as revealed in W-Ga. In fact, a structure of this type was recently reporrted for
nanocrystalline Al$_2$O$_3$/Ga composits \cite{Kon97}. It is expected that such skeleton structures are interesting materials
with respect to their magnetic behavior, i.e., magnetization, magneto-caloric effect or GMR.

\acknowledgements
The authors would like to thank H. Wolf and Th. Wichert for performing PAC measurements.

\begin{figure}
\caption{\label{material}
Schematics of a nanocrystalline material composed of a component A, representing the 
grains (open circles) and a component B, representing the grain boundary component (black 
circles).}
\end{figure}

\begin{figure}
\caption{\label{inert-gas}
Schematics of the vacuum chamber used for the inert gas condensation method, containing a 
tungsten sputter source and an evaporation source for Ga.}
\end{figure}

\begin{figure}
\caption{\label{XRD(W,W/Ga)}
(a) Diffraction pattern of a nanocrystalline W-A15 sample. The grain diameter is $\approx$ 13 nm. (b) 
Diffraction pattern of a nanocrystalline W-Ga sample. The Ga content is 40 at$\%$. All maxima correspond to 
 W-A15 with a grain diameter of $\approx$ 8 nm. No Bragg peaks corresponding to crystalline Ga could be detected.
}
\end{figure}

\begin{figure}
\caption{\label{DSC}
Thermal expansion during heating of a W-Ga composite (W grain diameter $\approx$ 10 nm, Ga content 
$\approx$ 30 at$\%$) (a) in comparison of a pure W sample (grain diameter $\approx$ 10 nm) (b) for 
temperatures between 130 and 350 K.}
\end{figure}

\begin{figure}
\caption{\label{DSC1}
(a) DSC signal of a W-Ga composite (30 at$\%$ Ga). The 
melting point of pure $\alpha$-Ga is 303 K. Neither at this temperature nor at the melting points 
of other modifications of Ga an endothermic signal is detected. (b) DSC signal of the same amount of pure Ga like in
(a).}
\end{figure}

\begin{figure}
\caption{\label{PAC}
(a) PAC spectrum of $^{111}$Cd in a nanocrystalline W-Ga composite (Ga content $\approx$ 30 at$\%$) at a 
temperature 280 K. (b) Frequency analysis of the PAC spectrum of (a). For comparison, the 
frequency analysis of amorphous Ga (shaded) according to Heubes et al. \cite{Heu79} (T $\leq$ 15 K) is 
shown.}
\end{figure}

\begin{figure}
\caption{\label{resistivity}
(a) Specific resistance of a nanocrystalline W-A15 sample as a function of the temperature. The grain diameter 
is $\approx$ 10 nm. (b) Specific resistance of a nanocrystalline W-Ga composite ($<$ 30 at$\%$ Ga) as a 
function of the temperature. The grain diameter of W is $\approx$ 10 nm.}
\end{figure}

\begin{figure}
\caption{\label{resistivity-ga}
(a) Model of a nanocrystalline W sample (left side) and of the Ga skeleton structure (right
side). (b) - (d) Resistivities of the W-Ga composite as a function of 
the temperature and varying Ga content. In the normal conductive state the TRC is a function 
of thickness ($\approx$ zero for about 40 at$\%$ Ga). The Ga content within the nanocrystalline
W samples is (b) W $<$ 30 at$\%$ Ga, (c) W/33 at$\%$ Ga, and (d) W/40 at$\%$ Ga.}
\end{figure}

\begin{figure}
\caption{\label{magnetic}
Magnetic behavior of a W-Ga composite (mean grain diameter of W $\approx$ 8 nm, Ga content 
$\approx$ 35 at$\%$) at T $<$ 5 K in a magnetic field.}
\end{figure}


\begin{references}
%
\bibitem{Gle81}
H. Gleiter, in: {\it Deformation of Polycrystals: Mechanism and Microstructures}, Proc. of the 
2nd Risø Int. Symp. on Metallurgy and Mat. Sci., M. Hansen, A. Horsewell, T. Leffers and H. 
Lilholt (eds.), p. 15, Verlag (1981).
%
\bibitem{Sie90}
R.W. Siegel, MRS Bulletin {\bf XV(10)}, 60 (1990).
%
\bibitem{Gle94}
H. Gleiter, {\it Nano'94 - Second International Conference on Nanostructured Materials}, 
Stuttgart (1990).
%
%
\bibitem{Jan90}
J.S.C. Jang and C.C. Koch, J. Mater. Res. {\bf 5}, 325 (1990).

\bibitem{Ogu78}
T. Ogura, C.J. McMahon, H.C. Feng, and V. Vitek, Acta Metall. {\bf 26}, 1317 (1978).

\bibitem{Hau92}
T. Haubold, F. Boscherini, S. Pascarelli, S. Mobilio, and H. Gleiter, Phil. Mag. {\bf 66}, 591 (1992).

\bibitem{Tur90}
D. Turnbull, J.S.C. Jang, and C.C. Koch, J. Mater. Res. {\bf 5}, 1731 (1990).

\bibitem{Yan93}
M. Yan, M. Sob, D.E. Luzzi, V. Vitek, G.J. Ackland, M. Methfessel, and C.O. Rodriguez, Phys. Rev. {\bf
B47}, 5571 (1993).

\bibitem{Boo92}
P. Boolchand and C.C. Koch, J. Mater. Res. {\bf 7}, 2876 (1992).

\bibitem{Kon97}
H. Konrad, C. Karmonik, J. Weissm\"uller, H. Gleiter, R. Birringer, and R. Hempelmann, Physica {\bf
B234-236}, 173 (1997).

\bibitem{Hor98}
J.J. Horst, J.A. Block, K. Parvin, V.P. Dravid, J.L. Alpers, T. Sezen, and R. La
Duca,
 J. Appl. Phys. {\bf83}, 793 (1998).
 
\bibitem{Ell98} 
 B.R. Elliot, J.J. Horst, V.P. Dravid, M.H. Teng, and J.H. Hweng, J. Mat. Res.
 {\bf
12}, 3328 (1998).

\bibitem{Nis73}
O. Nishikawa and T. Utsumi, J. Appl. Phys. {\bf 44}, 955 (1973).

\bibitem{Kon81}
M. Konishi, M. Wada and O. Nishikawa, Surf. Sci. {\bf 108}, L463 (1981).

\bibitem{Mas86}
T.B. Massalski (ed.), {\it Binary Alloy Phase Diagrams}, American Society for Metals (Metals
Park, Ohio, 1986).

\bibitem{Hah90}
H. Hahn and R.S. Averback, J. Appl. Phys. {\bf 67}, 1113 (1990).

\bibitem{Haa92}
V. Haas and R. Birringer, Nanostruct. Mat. {\bf 11}, 491 (1992).

\bibitem{Wei93}
J. Weissm\"uller, Nanostruct. Mater. {\bf 3}, 261 (1993).

\bibitem{Kra97}
W. Krauss and R. Birringer, Nanostruct. Mat. {\bf 9}, 109 (1997).

\bibitem{Cor93}
L. Cortella, B. Vinet, P.J. Desr\'e, A. Pasturel, A.T. Paxton, and M. van Schilfgaarde, Phys. 
Rev. Lett. {\bf 70}, 1469 (1993).

\bibitem{Kli92}
M. Klingel, PhD thesis, University of Saarland (1992).

\bibitem{Heu79}
P. Heubes, D. Korn, G. Schatz and G. Ziebold, Phys. Lett. {\bf 74A}, 267 (1979).

\bibitem{Wol94}
H. Wolf, H.G. Zimmer, T. Filz and Th Wichert, {\it Nano94 - Second International 
Conference on Nanostructured Materials} (Stuttgart, 1994).

\bibitem{Ges72}
E.I. Geshko, V.P. Mikhal'chenko and B.M. Sharlai, Sov. Phys. Solid State {\bf 14}, 1554 
(1972).

\bibitem{Yaq66}
M. Yaqub, D. Waldorf, R. Boughton and W.A. Jeffers, Phys. Lett. {\bf 23}, 423 (1966).

\bibitem{Buc63}
W. Buckel and W. Gey, Z. Phys. {\bf 176}, 337 (1963).

\bibitem{Gin86}
G. Ginter, J.G. Gasser and R. Kleim, Phil. Mag. {\bf B54}, 543 (1986).

\bibitem{Kir80}
Kirk-Othmer, {\it Encyclopedia of Cemical Technology} {\bf 11} (3rd ed.) (1980), p. 
606.

\bibitem{Buc84}
W. Buckel, {\it Supraleitung} (Physik-Verlag, 1984), p.164.

\bibitem{Cam64}
A. Campell, J.E. Evetts and D. DewHughes, Phil. Mag. {\bf 10}, 333 (1964).


\bibitem{Coc65}
J.F. Cochran and M. Yaqub, Phys. Rev. {\bf 140A}, 2174 (1965).


\end{references}
\end{document}